\documentclass[aps,amsmath,showpacs]{revtex4}
\usepackage{amsmath}
\usepackage{graphicx}
\begin{document}
\title{Nonlinear stationary solutions of the Wig\-ner and Wigner-Poisson equations}
\author{F. Haas \footnote{Also at Universidade do Vale do Rio dos Sinos - UNISINOS, Av. Unisinos 950, 93022\---000, S\~ao Leopoldo, RS, Brazil} \,and P. K. Shukla \footnote{Also at Department of Physics, Ume\aa \, University, SE-90187, Ume\aa, Sweden; GOLP / Instituto de Plasmas e Fus\~ao Nuclear, Instituto Superior T\'ecnico, Universidade T\'ecnica de Lisboa, 1049-001 Lisboa, Portugal; SUPA, Department of Physics, University of Strathclyde, Glasgow, G40NG, UK; School of Physics, University of Kwazulu-Natal, Durban 4000, South Africa.}}
\revised{16 September 2008}
\affiliation{Institut f\"ur Theoretische Physik IV, Ruhr--Universit\"at Bochum,
D-44780 Bochum, Germany}

\begin{abstract}
\noindent
Exact nonlinear stationary solutions of the one-dimensional Wigner and Wigner-Poisson equations in the terms of the Wigner functions that depend not only on the energy but also on position are presented. In this way, the Bernstein-Greene-Kruskal modes of the classical plasma are adapted for the quantum formalism in the phase space. The solutions are constructed for the case of a quartic oscillator potential, as well as for the self-consistent Wigner-Poisson case. Conditions for well-behaved physically meaningful equilibrium Wigner functions are discussed. 
\end{abstract}


\maketitle

\section{Introduction}
It is well-known that the solutions for the Vlasov equation are ar\-bi\-tra\-ry func\-tions of the invariants (constants of motion) of the system. In the sta\-tio\-na\-ry case, this property allows us to construct the so-called Bern\-stein-Gree\-ne-Krus\-kal equilibria \cite{bgk} for the Vlasov-Poisson system for the classical plasma. On the other hand, it is much more difficult to derive exact solutions for the corres\-pon\-ding quantum model, namely the Wigner equation or, in the self-con\-sis\-tent case, the Wigner-Poisson system. Indeed, even if the classical problem is integrable, the stationary Wigner function is not, in general, a function of the classical invariants. This follows since the Wigner time-evolution equation does not preserve the classical constants of motion. Therefore, to date, there is a lack of exact solutions for the Wigner and Wigner-Poisson equations. The better results in this regard are approximate solutions \cite{luque}--\cite{smerzi} that have been obtained as the first order quantum correction to the Vlasov-Poisson equilibria. The quantum Bernstein-Greene-Kruskal modes have been defined as the solutions of the Wigner-Poisson system under the periodic boundary conditions \cite{lange}, but the explicit construction of these solutions is still a challenge. Notice that the original article by Wigner \cite{wig}, where he introduces his celebrated function,  considers the first-order quantum correction, to a Maxwell-Boltzmann thermodynamic equilibrium. In addition, quantum like corrections were proposed for charged-particle beam transport \cite{fedele}.  

The purpose of the present paper is to develop explicit exact solutions of the Wigner equation, in a particular form that is described in Section II. Both the external and self-consistent potential cases are treated. In a sense, we develop a quantum analogue of the 
Bernstein-Greene-Kruskal equilibria in the phase space. Of course, exact stationary Wigner functions can be found after applying the Wigner transform to the previously known energy eigenstates, if available. However, the question is to derive solutions for the Wigner equation considered in itself, without reference to the Schr\"odinger equation or the nature of the quantum statistical mixture of the system. Here we propose a special functional {\it Ansatz}, as a function of the energy and the position, for the Wigner function in a conservative system. The technique is illustrated for a quartic anharmonic oscillator case and for the self-consistent Wigner-Poisson system. The derived solutions are the first explicit exact solutions of the Wigner equation for nonlinear systems, constructed independently of the Schr\"odinger formalism. In this way, we can have insight on the relation between the classical constants of motion and the solutions of the sta\-tio\-na\-ry Wigner equation. In this regard, it is already well known \cite{schuch} that for a quadratic Hamiltonian there is a close relationship between the Er\-ma\-kov-Le\-wis invariant, which is the basic classical constant of motion of the problem, and the Wigner function. However, for a quadratic Hamiltonian the Wigner equation reduces to the classical (Vlasov) equation, with the quantum effect restricted to the initial conditions. 

The work is organized as follows. In Section II, we consider a quartic nonlinear oscillator problem and derive quantum phase space structures by using our {\it Ansatz} generalizing the Bernstein-Greene-Kruskal solutions for the classical system. In Section III, the approach is adapted to the Wigner-Poisson system, which is related to a quantum plasma system. Section IV is reserved to the conclusions. 

\section{Exact solutions of the stationary Wigner equation for a nonlinear oscillator potential}
The Wigner equation \cite{carru} in one spatial dimension with a potential $V(x,t)$ reads 
\begin{eqnarray}
\label{eq1}
\frac{\partial\,f}{\partial\,t} &+& v\frac{\partial\,f}{\partial\,x}   \\  &=& \frac{im}{2\pi\hbar}\int\,d\lambda\,dv'\,e^{im(v-v')\lambda}\left[V(x+\frac{\lambda\hbar}{2},t)-V(x-\frac{\lambda\hbar}{2},t)\right]\,f(x,v',t) \,,\nonumber
\end{eqnarray}
where $f(x,v,t)$ is the Wigner pseudo-probability distribution and all quantities have their usual meaning. The formal classical limit ($\hbar \rightarrow 0$) of the Wigner equation is the Vlasov equation, with the Wigner function playing the role of the probability distribution function. Even assuming negative values in some regions of phase space, the Wigner function can be used to compute the macroscopic quantities like the density and the  current, in the same way as a faithful distribution function.  In addition to Eq. (\ref{eq1}), a genuine Wigner function should correspond to a positive definite density matrix. Therefore, $f = f(x,v,t)$ must satisfy \cite{hillery} at least the following necessary conditions, 
\begin{eqnarray}
\label{c1}
\int dx dv f &=& 1 \,,\\
\label{c3}
\int dv f &\geq& 0 \,,\\
\label{c4}
\int dx f &\geq& 0 \,,\\
\label{c2}
\int dx dv f^2 &\leq& \frac{m}{2\pi\hbar} \,.
\end{eqnarray}
Equation (\ref{c1}) is just a normalization condition, while Eqs. (\ref{c3}) and (\ref{c4}) arise because the spatial and velocity densities should be  non-negative everywhere. Finally, Eq. (\ref{c2}) is needed to avoid violation of the uncertainty principle, ruling out a too spiky function $f(x,v,t)$.   

In the stationary case, one would be tempted to use the conservation of the energy to construct exact solutions for Eq. (\ref{eq1}). However, except in the somewhat restricted case where the Wigner function is a linear function of the energy \cite{manfeix}, the Wigner equation is not satisfied by functions of the energy alone. To illustrate our technique, we consider the anharmonic potential 
\begin{equation}
\label{eq2}
V = \mu \Bigl(\frac{m\omega^2 x^2}{2} - \frac{m\omega^2 k^2 x^4}{24}\Bigr) \,,
\end{equation}
where $\mu = \pm 1$ is a numerical parameter and $\omega$ and $k$ are parameters with dimension of angular frequency and wavenumber, respectively. Plots of the potential are shown in Figure 1, showing a single nonlinear potential well or a double-well potential, according to the values of $\mu$. The coefficients were chosen to match the expansion of a pendulum-like potential $\mu m(\omega^{2}/k^{2})[1 - \cos(kx)]$. A similar form was considered in connection with the numerical simulation of quantum echoes described by the Wigner equation \cite{echo}.

\begin{figure*}
\begin{center}
\includegraphics{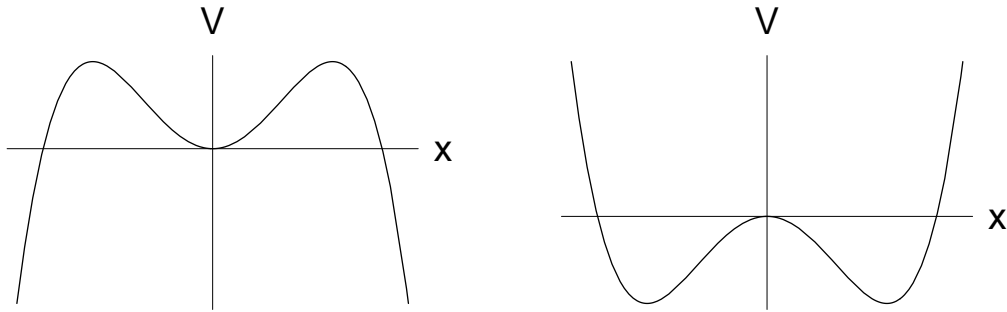}
\caption{the potential in Eq. (\ref{eq2}) for $\mu = 1$ and $\mu = -1$, respectively.}
\label{figure1}
\end{center}
\end{figure*}

For the chosen potential, in the stationary case ($\partial f/\partial t = 0$), the Wigner equation (\ref{eq1}) is 
\begin{equation}
\label{eq3}
v\frac{\partial f}{\partial x} - \mu(\omega^2 x - \frac{\omega^2 k^2 x^3}{6})\frac{\partial f}{\partial v} = \frac{\mu \hbar^2 \omega^2 k^2 x}{6m^2} \frac{\partial^3 f}{\partial v^3} \,.
\end{equation}
If there was no nonlinearity,  Eq. (\ref{eq3})  would be the stationary Vlasov equation for the simple harmonic oscillator whose solutions are arbitrary functions of the energy. However, in the quantum and nonlinear case, the higher-order velocity derivative term prevents the existence of a solution as a function of the energy only. Notice that the term in the right-hand side of Eq. (\ref{eq3}) can assume large values: it is not necessarily just a quantum correction. Higher-order (in $\hbar^2$) terms would appear for higher anharmonicities. 

With the rescaling $q = kx,  p = kv/\omega, F = \omega f/k^2$, Eq. (\ref{eq3}) transforms into  
\begin{equation}
\label{eq4}
p\frac{\partial F}{\partial q} - \mu (q - \frac{q^3}{6})\frac{\partial F}{\partial p} = \mu \Gamma q \frac{\partial^3 F}{\partial p^3} \,,
\end{equation}
where
\begin{equation}
\label{eq5}
\Gamma = \frac{\hbar^2 k^4}{6m^2 \omega^2} 
\end{equation}
is a non-dimensional parameter measuring the relevance of the quantum effect. At this point, notice that the necessary conditions (\ref{c1})--(\ref{c2}) are rewritten as 
\begin{eqnarray}
\label{d1}
\int dq dp F &=& 1 \,,\\
\label{d3}
\int dp F &\geq& 0 \,,\\
\label{d4}
\int dq F &\geq& 0 \,, \\
\label{d2}
\int dq dp F^2 &\leq& \frac{1}{2\pi\sqrt{6\Gamma}} \,.
\end{eqnarray}

It is more convenient  to search for solutions of Eq. (\ref{eq4}) in the form $F = F(H,q)$, where  
\begin{equation}
\label{eq6}
H = \frac{p^2}{2} + \mu\left(\frac{q^2}{2} - \frac{q^4}{24}\right) 
\end{equation}
is the energy.  Then from Eq. (\ref{eq4}) we obtain 
\begin{equation}
\label{eq7}
\frac{\partial F}{\partial q} = \mu \Gamma q \left(\Bigl[2H - \mu (q^2 - \frac{q^4}{12})\Bigr] \frac{\partial^3 F}{\partial H^3} + 3 \frac{\partial^2 F}{\partial H^2}\right) \,.
\end{equation}
The usefulness of the energy and position variables in the treatment of the Wigner equation has been recognized  in the semi-classic case \cite{demeio}, but we feel that this method can be pursued in more depth for a fully quantum system. 
When there are no quantum effects, the right-hand side of Eq. (\ref{eq7}) is zero and the solution is simply $F = F(H)$, with an arbitrary functional dependence, in the same spirit of the Bernstein-Greene-Kruskal solutions for the classical plasma. The difficulties with Eq. (\ref{eq7}) are in the entangled character of the quantum term, which prevents the use of the separation of variables technique, for instance. It is also apparent that if $\partial F/\partial q \equiv 0$, then any linear function of the energy would be a solution. 

Our proposal is to consider exact solutions according to 
\begin{equation}
\label{eq8}
F = [A(q)H + B(q)] e^{C(q)H} \,,
\end{equation}
where $A, B$ and $C$ are functions to be determined, depending on position only. For $A = 0$, the Wigner function automatically has a Gaussian shape in velocity space. Otherwise, for $A \neq 0$, one can have a 
two-stream type Wigner function, double-humped in velocity space, at least for specific parameters. Notice that the proposed solution does not distinguish between the trapped and untrapped particles. This is another different feature in comparison to the Bernstein-Greene-Kruskal method. 

After inserting Eq. (\ref{eq8}) into Eq. (\ref{eq7}), the exponential factorizes and we derive a second-degree polynomial of $H$, which must be identically zero. Setting the coefficients of the different powers of the energy to zero, we obtain  
\begin{eqnarray}
A' + BC' &=& \mu \Gamma q \Bigl[9AC^2 + 2BC^3 - \mu AC^3 (q^2 - \frac{q^4}{12})\Bigr] \,, \nonumber \\
B' &=& \mu \Gamma q \Bigl[6AC + 3BC^2 - \mu(3AC^2 + BC^3) (q^2 - \frac{q^4}{12})\Bigr] \,, \label{eq9}  \\
AC' &=& 2\mu \Gamma q C^3 A \,, \nonumber
\end{eqnarray}
where the prime denotes derivative with respect to $q$.

For $B = 0$, it can be verified that the above system do not admit any solutions. For $B \neq 0$, the result is
\begin{eqnarray}
A &=& A_0 (C_0 - 2\mu\Gamma q^2)^{-9/4} \times \nonumber \\ \label{eq11} &\times& \exp\Bigl\{\frac{2C_{0}^2 - 2\mu\Gamma C_0 (18 +  q^2) +  \Gamma^2 q^2 (36 -  q^2)}{72\mu \Gamma^2 (C_0 - 2\mu \Gamma q^2)^{1/2}}\Bigr\} \,, \\
B &=& \Bigl\{B_0 (C_0 - 2\mu \Gamma q^2)^{-3/4} + A_0 \Bigl[\frac{3}{2 (C_0 - 2\mu \Gamma q^2)^{7/4}} + \nonumber \\ &+& \frac{\mu}{24\Gamma^2 (C_0 - 2\mu \Gamma q^2)^{9/4}} (- 2 C_{0}^2 + 6 \mu C_0 \Gamma (2 +  q^2) - 3\Gamma^2 q^2 (q^2 + 12))\Bigr]
\Bigr\} \times \nonumber \\ \label{eq12} &\times&
\exp\Bigl\{\frac{2C_{0}^2 - 2\mu\Gamma C_0 (18 +  q^2) +  \Gamma^2 q^2 (36 -  q^2)}{72\mu \Gamma^2 (C_0 - 2\mu \Gamma q^2)^{1/2}}\Bigr\}  \,,\\
\label{eq10}
C &=& -  \frac{1}{(C_0 - 2\mu\Gamma q^2)^{1/2}} \,,  
\end{eqnarray}
where $A_0, B_0$ and $C_0 \neq 0$ are integration constants. If $A_0 \equiv 0$, then the Wigner function is certainly Gaussian in velocity space. 

In the double-well potential case, when $\mu = - 1$, the exact solution is real and bounded provided $C_0 > 0$. However, for $\mu = 1$, it can be shown that the exact solution becomes singular  when $q^2 \rightarrow C_{0}/(2\Gamma)$, unless $0 < C_0 < 24\Gamma$. Actually, when $\mu = 1$ the Wigner function in Eq. (\ref{eq8}) is acceptable only for $q^2 \leq C_{0}/(2\Gamma)$. Figure 2 displays a typical plot in phase space, for $\Gamma = 1, \mu = 1, A_0 = 0, B_0 = 0.42, C_0 = 16$. For simplicity, we assume that there are no scattering states, so that $F \equiv 0$ for $q^2 > C_{0}/(2\Gamma) = 8$. The Wigner function has a Gaussian shape in the momentum space and an abrupt localization in configuration space, as can be seen more clearly in Figure 3 for the same parameters and for different rescaled velocities $p = 0$ and $p = 2$. These strongly localized structures do not  have classical counterpart. Observe that the necessary conditions in Eqs. (\ref{d1})-(\ref{d2}) are also satisfied. In particular, $\int dq dp F^2 = 0.03 < (2\pi\sqrt{6\Gamma})^{-1} = 0.07$. 

\begin{figure*}
\begin{center}
\includegraphics{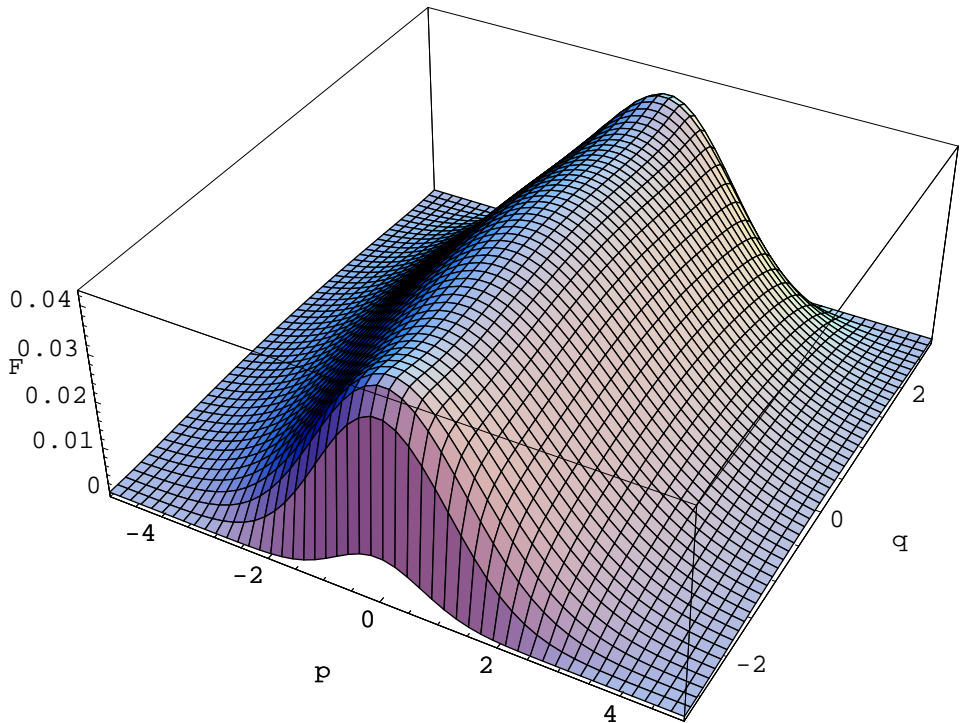}
\caption{the exact Wigner function in Eq. (\ref{eq8}) for $\Gamma = 1, \mu = 1, A_0 = 0, B_0 = 0.42, C_0 = 16$.}
\label{figure2}
\end{center}
\end{figure*}

\begin{figure*}
\begin{center}
\includegraphics{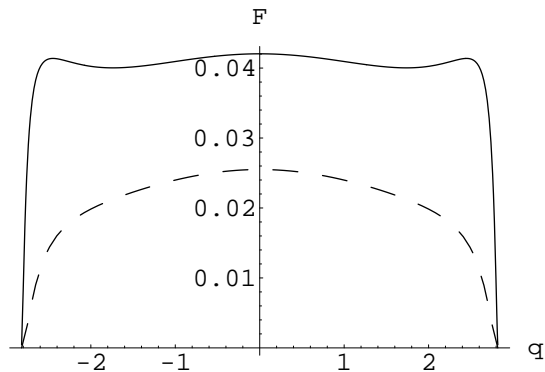}
\caption{spatial localization of the exact Wigner function in Eq. (\ref{eq8}) for $\Gamma = 1, \mu = 1, A_0 = 0, B_0 = 0.42, C_0 = 16$ and $p = 0$ (solid line) and $p = 2$ (dashed line).}
\label{figure3}
\end{center}
\end{figure*}

When $\mu = -1$, the Wigner function is regular provided $C_0 > 0$. Then, we can have a more rich variety of behaviors. For instance, for $\mu = -1, A_0 > 0, B_0 = -2.5 A_{0}, C_0 = 1$, there is a two-humped distribution in velocity space, provided $\Gamma > 0.37$.  Figure 4 shows the Wigner function in momentum space, for fixed position $q = 0$, with the same parameters and several values of $\Gamma$. One sees that for increasing quantum effects there is a progressive depth of the exact solution, which eventually becomes negative for $\Gamma > 0.63$. Notice that, in principle, there is no limiting value of $\Gamma$, since the solution is non perturbative. However, for a physically meaningful positive particle spatial density
\begin{equation}
\label{d}
n(q) = \int dp F = \frac{\sqrt{\pi}}{\sqrt{2|C|}}\exp\left[-\frac{\mu|C|}{2}(q^2 - \frac{q^4}{12})\right] \left(\frac{A}{|C|} +  \mu A (q^2 - \frac{q^4}{12}) + 2B\right) 
\end{equation}
there are extra restrictions, even recognizing that in general the Wigner function is not a positive definite quantity. For instance, when $\mu = -1, A_0 > 0, B_0 = - 2.5 A_{0}, C_0 = 1$, $n(q) \geq 0$ for all $q$ imposes that $\Gamma \leq 0.3$. For these parameters, the other necessary conditions in Eqs. (\ref{d1}), (\ref{d4}) and (\ref{d2}) are also fulfilled, as can be checked after numerically performing the spatial integrations. Actually, a detailed calculation shows that when $\mu = -1$, $A_0 > 0$ and $B_0 < 0$, for a positive definite particle density the exact Wigner function cannot be a two-stream type distribution. On the other hand, for $\mu = -1$ and non negative $A_0, B_0$ (in which case the distribution function is not two-humped in momentum space), in principle one can have arbitrarily large values of $\Gamma$.  

\begin{figure*}
\begin{center}
\includegraphics{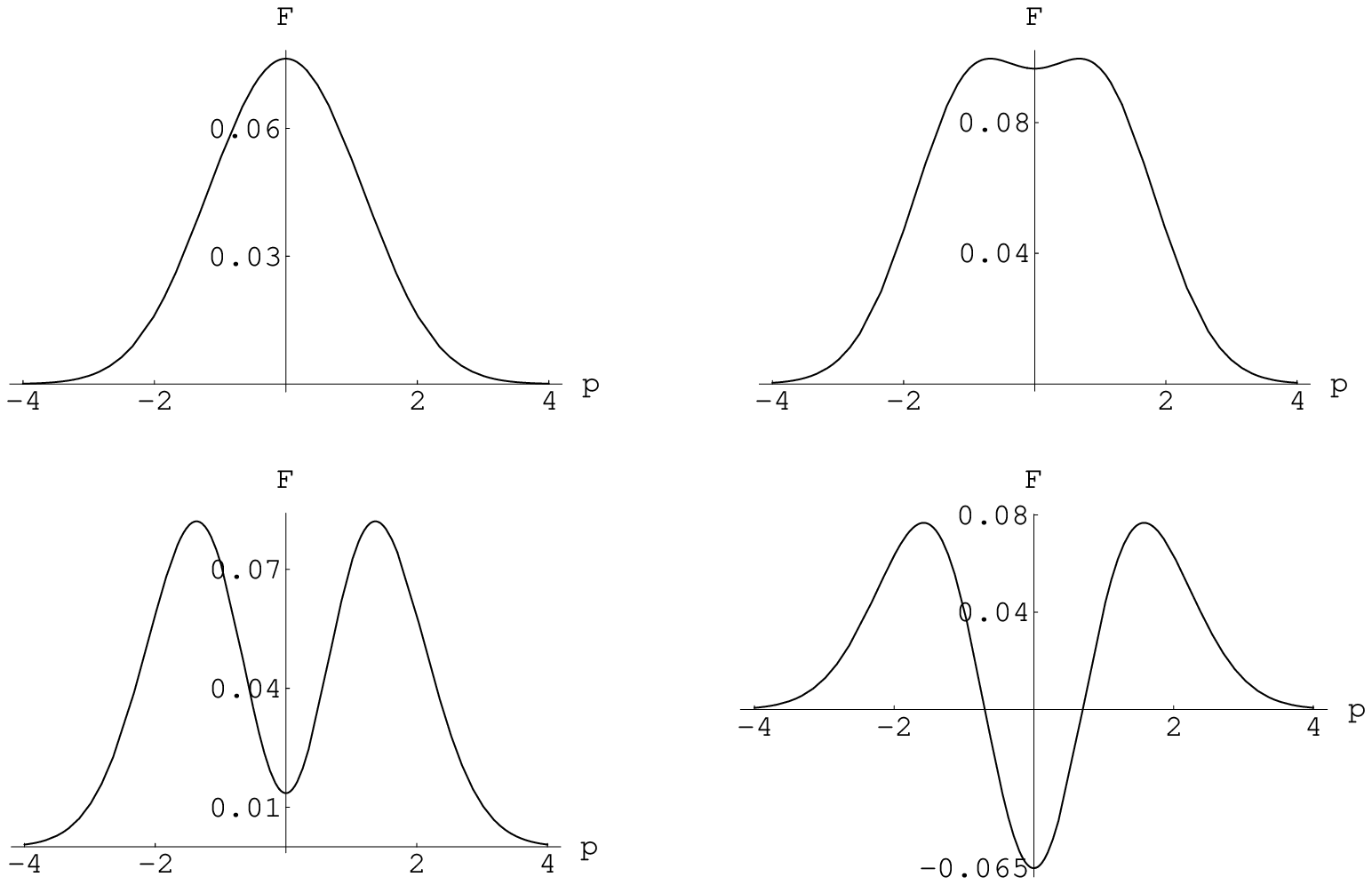}
\caption{the solution in Eq. (\ref{eq8}) in momentum space for $q = 0$, $\mu = - 1$, $A_0 = 0.52$, $B_0 = - 1.3$ and $C_0 = 1$. The quantum parameter is $\Gamma = 0.2$ (left, top), $\Gamma = 0.4$ (right, top), $\Gamma = 0.6$ (left, bottom) and $\Gamma = 0.8$ (right, bottom). In the graphs, only for $\Gamma = 0.2$ the solution is a physically acceptable Wigner function.}
\label{figure4}
\end{center}
\end{figure*}

The opposite behavior arises when $\mu = - 1$ and $A_0 < 0, B_0 > 0$. In this case, it can be proved that a positive definite particle density implies a two-humped Wigner function in momentum space. For instance, at the origin $q = 0$, the condition for $n(0) \geq 0$ reads
\begin{equation}
\label{c}
12 B_0 C_{0}^{3/2} \Gamma^2 \geq |A_0| \left(C_{0}^2 + 6\Gamma C_0 + 24 \Gamma^2 \sqrt{C_0}\right) \,,
\end{equation}
and this inequality can be shown to imply that the Wigner function admits two minima in velocity space, as illustrated in Figure 5. Notice the opposite concavity in comparison to the $\mu = -1, A_0 > 0, B_0 < 0$ case. In the same figure, we show the particle density, which has a deep centered at the origin but is positive definite, even with  the Wigner function admitting negative values. 

Now, it can happen that the quantum effect comes in favor of a physically meaningful solution, since $n(q) \geq 0$ for all $q$ provided $\Gamma$ is sufficiently large. For instance,  from Eq. (\ref{c}) it follows that $n(0) \geq 0$ if $B_0 C_0 > 2 |A_0|$ and $\Gamma$ is large enough. In particular, for $\mu = -1, A_0 < 0, B_0 = - 2.5 A_{0}$ and $C_0 = 1$, one needs $\Gamma \geq 1.14$. Figure 6 shows $n(q)$ for these parameters and several values of $\Gamma$. One sees that the two maxima of the particle density tend to approximate each other as the quantum parameter increases. Also, the minimum and maximum values of $n(q)$ increases as quantum effects becomes larger. However, $\Gamma$ can't be arbitrarily large, since the solution becomes too localized in phase space for increasing $\Gamma$. For instance, when $\Gamma = 5, \mu = -1, A_0 = -0.11, B_0 = 0.28$ (figure 6, right, bottom), we have $\int dq dp F^2 = 0.032 > (2\pi\sqrt{6\Gamma})^{-1} = 0.029$, violating Eq. (\ref{d2}).  In conclusion, in each specific case the necessary conditions (\ref{d1})--(\ref{d2}) must be checked. 

\begin{figure*}
\begin{center}
\includegraphics{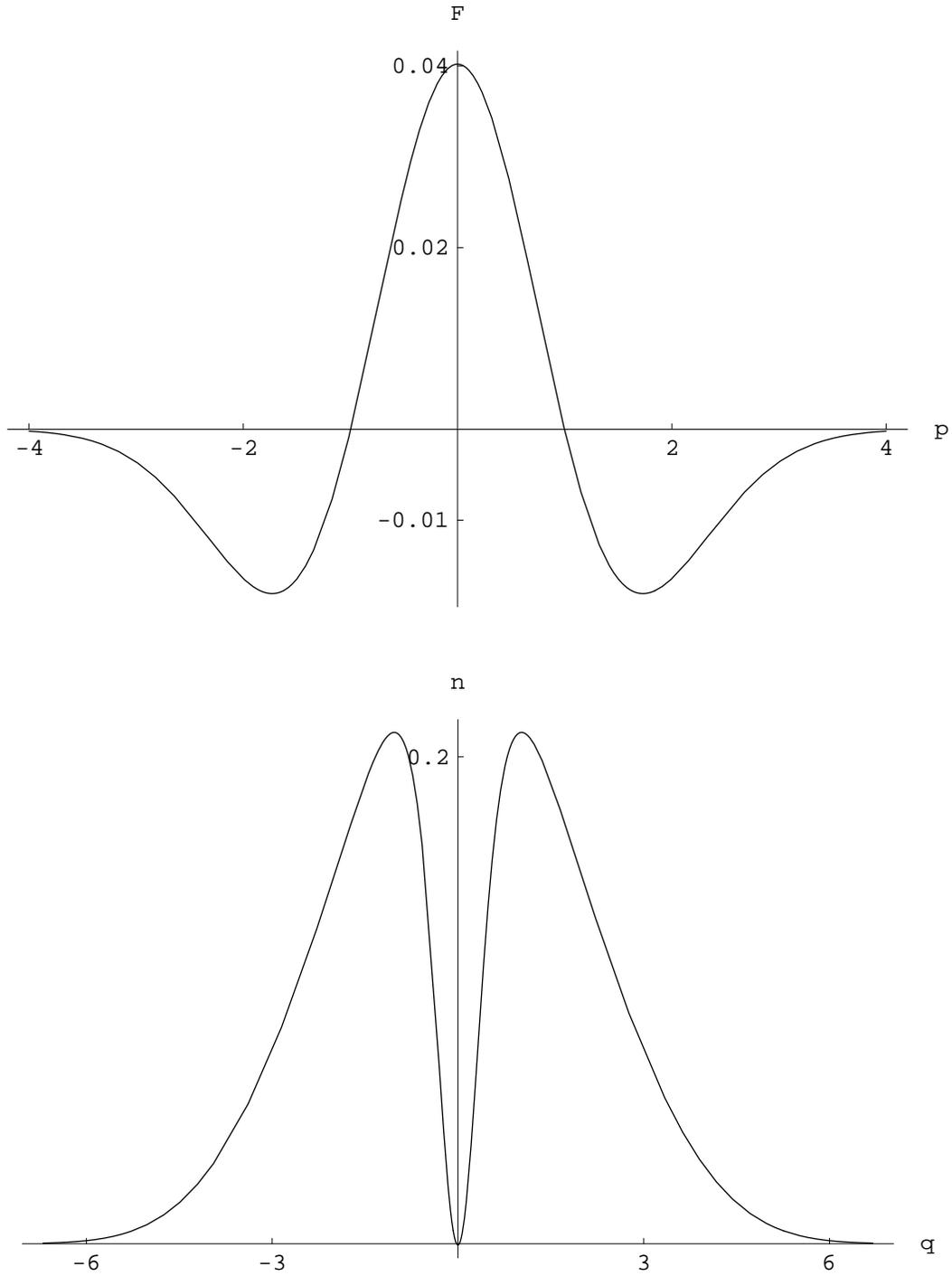}
\caption{on the top: the Wigner function in Eq. (\ref{eq8}) for $q = 0, \mu = -1, \Gamma = 1.14,  A_0 = - 0.13, B_0 = 0.32, C_0 = 1$. On the bottom: the corresponding particle number density.}
\label{figure5}
\end{center}
\end{figure*}

\begin{figure*}
\begin{center}
\includegraphics{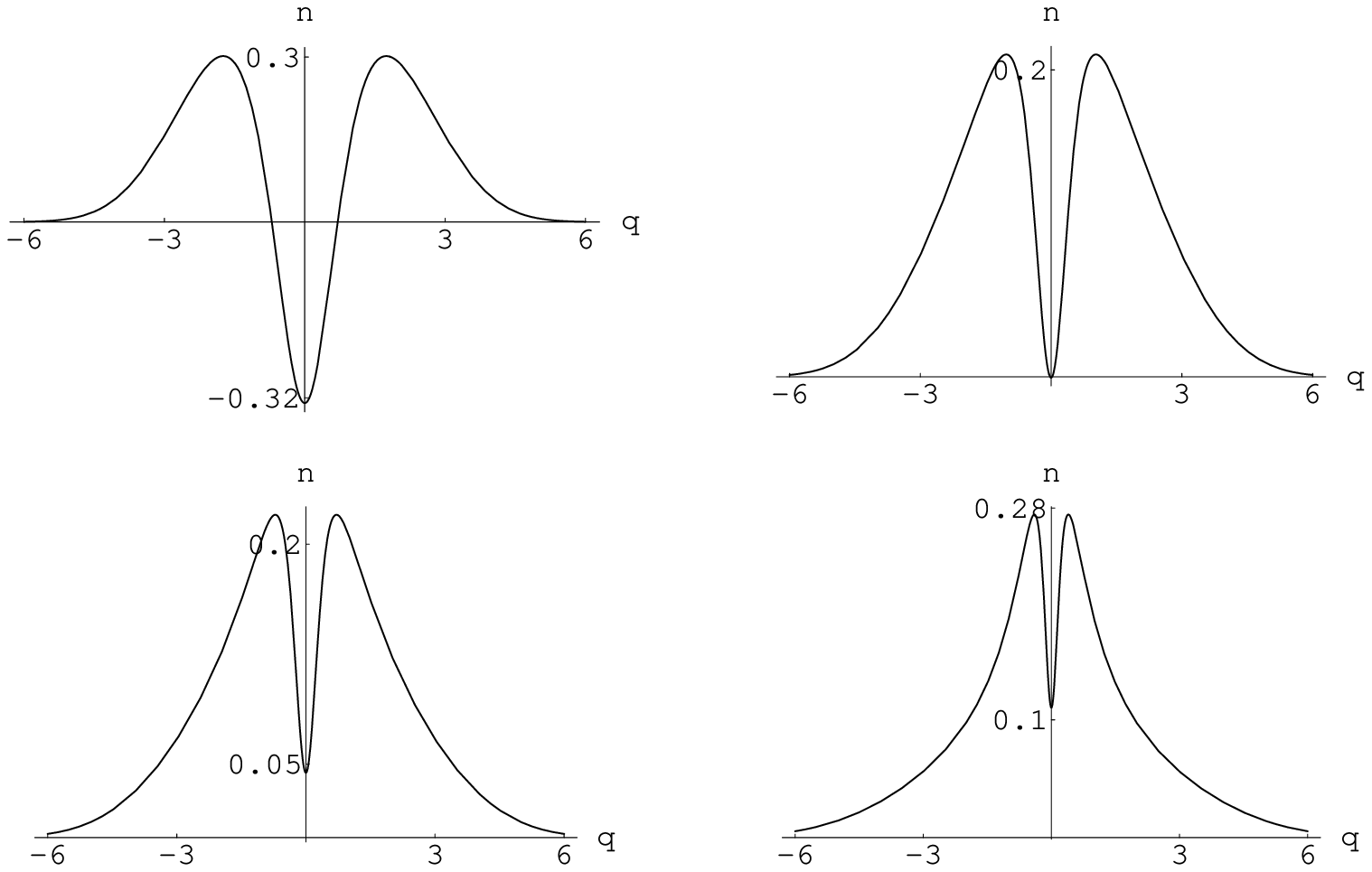}
\caption{particle number density in Eq. (\ref{d}) for $\mu = -1,  B_0 = -2.5 A_{0}, C_0 = 1$ and for $\Gamma = 0.5$, $A_0 = -0.48$ (left, top),  $\Gamma = 1.14$, $A_0 = -0.13$ (right, top), $\Gamma = 2$, $A_0 = -0.108$ (left, bottom) and  $\Gamma = 5$, $A_0 = - 0.111$ (right, bottom). For each $\Gamma$, a different $A_0$ must be chosen, to comply with a normalized Wigner function. The derived solution complies with the conditions (\ref{d1})--(\ref{d2}), with the exception of the case $\Gamma = 5$, $A_0 = - 0.111$.}
\label{figure6}
\end{center}
\end{figure*}

\section{Application to the Wigner-Poisson system}
Now consider the self-consistent case where the potential is $V = - e\phi$, where $e$ is the elementary charge and $\phi$ is the electrostatic potential satisfying the Poisson equation 
\begin{equation}
\label{eq13}
\frac{\partial^2 \phi}{\partial x^2} = \frac{e}{\epsilon_0} (n - n_{0}) \,, \quad n = n(\phi) = \int f dv \,,
\end{equation}
where $\epsilon_0$ is the permittivity constant and $n_0$ a background ion number density. Coupled to the Wigner equation, Eq. (\ref{eq13}) compose the Wigner-Poisson system \cite{new}. For the latter, it is convenient to search for solutions in the form $f = f(H,n)$, where $H = mv^{2}/2 - e\phi$ is the energy and $n$ is the electron number density, as defined in Eq. (\ref{eq13}). Then, writing the stationary Wigner equation, retaining only the first-order quantum correction, 
\begin{equation}
\label{wpq}
v\frac{\partial f}{\partial x} + \frac{e}{m}\frac{\partial\phi}{\partial x}\frac{\partial f}{\partial v} - \frac{e\hbar^2}{24m^3}\frac{\partial^3 \phi}{\partial x^3} \frac{\partial^3 f}{\partial v^3} = 0 \,,
\end{equation}
we obtain, after using the Poisson equation, 
\begin{equation}
\label{eq14}
\frac{\partial f}{\partial n} = \frac{\hbar^2 \omega_{p}^2}{24n_0} \left[2\Bigl(H + e\phi(n)\Bigr)\frac{\partial^3 f}{\partial H^3} + 3\frac{\partial^2 f}{\partial H^2}\right] \,,
\end{equation}
where $\omega_p = (n_0 e^{2}/m\epsilon_0)^{1/2}$ is the electron plasma frequency. In Eq. (\ref{eq14}), the scalar potential is interpreted as a function of the electron number density. This is locally possible using $n = n(\phi)$, according to the implicit function theorem, provided $n$ is not identically a constant. Now the {\it Ansatz}
\begin{equation}
\label{eq15}
f = [A(n)H + B(n)]e^{C(n)H}
\end{equation}
produces the following system for the functions $A, B$ and $C$ depending on the electron number density only, 
\begin{eqnarray} 
A' + BC' &=& \frac{\hbar^2 \omega_{p}^2}{24n_0} \Bigl[2 (B + e\phi A) C + 9A\Bigr] C^2 \,, \nonumber \\ \label{eq16} 
B' &=& \frac{\hbar^2 \omega_{p}^2}{24n_0} \Bigl[2e\phi C^2 (BC + 3A) + 3C (BC + 2A)\Bigr] \,,  \\
AC' &=& \frac{\hbar^2 \omega_{p}^2}{12n_0} AC^3 \,, \nonumber
\end{eqnarray}
where the prime denotes derivative with respect to $n$. The resulting system can be solved in all generality, but here we are content with the $A \equiv 0$ case, which is more amenable to detailed calculations. When $A = 0$, the solution for the system (\ref{eq16}) is given recursively as
\begin{eqnarray}
\label{eq17}
C&=& - \frac{C_{0}}{(1 - \hbar^2 \omega_{p}^2 C_{0}^2 n/(6n_{0}))^{1/2}} \,, \\ \label{eq18}
B&=&B_{0}\left(1-\frac{\hbar^{2}\omega_{p}^{2}C_{0}^{2}n}{6n_0}\right)^{-3/4}\exp\left[\frac{e\hbar^{2}\omega_{p}^2}{12n_0}\int\,dn\phi(n)C^{3}(n)\right]\,,
\end{eqnarray}
where $B_0$ and $C_0$ are integration constants, with $C_0 > 0$ for an integrable Wigner function.

Using Eq. (\ref{eq15}), we obtain  
\begin{equation}
\label{eq19}
n(\phi) = B \sqrt{\frac{2\pi}{-Cm}} \exp(-eC\phi)  \,.
\end{equation}
After rearranging, Eq. (\ref{eq18}) then gives the following integral equation for $B = B(n)$, 
\begin{equation}
\label{eq20}
B = B_0 \left(1 - \frac{\hbar^2 \omega_{p}^2 C_{0}^2 n}{6n_0}\right)^{-3/4} \exp\Bigl[- \frac{\hbar^2 \omega_{p}^2 }{12n_0} \int dn C^{2}(n) \ln\left(\frac{n}{B}\sqrt{\frac{-C(n) m}{2\pi}}\right)\Bigr] \,,
\end{equation}
where $C = C(n)$ is given by Eq. (\ref{eq17}). 

In the absence of the quantum effect, from Eq. (\ref{eq18}), one would directly have $B = B_0$. Since in the self-consistent case we are considering, for simplicity, only the first-order quantum correction, it is then reasonable to put $B = B_0$ in the integral at the right-hand side of Eq. (\ref{eq20}). After performing the integral, which is a complicated expression involving the dilogarithm function ${\rm Li}_{2}(z) = \int_{1}^{z} ds \ln(s)/(1-s)$, one derives  a function $B = B(n)$ and hence the electrostatic potential $\phi = \phi(n)$ through Eq. (\ref{eq19}). As apparent from Eqs. (\ref{eq17}), (\ref{eq18}) and (\ref{eq20}), one can identify the non-dimensional quantum parameter $\hbar^2 \omega_{p}^2 C_{0}^2/6 $, with the interpretation of the inverse square of a temperature.  Figure 7 displays typical plots of $\phi(n)$, with units so that $e =m = \hbar = n_0 = \omega_p  = B_0 = 1$, for different values of $C_0$. The graphs are similar to that from a electron hole Maxwell-Boltzmann equilibrium ($\phi \sim \ln n$), however with modifications and a cutoff when $n \rightarrow 6n_{0}/(\hbar^2 \omega_{p}^2 C_{0}^2)$. For increasing quantum effects (larger $C_0$, or a smaller effective temperature), the cutoff becomes closer to the origin. The explicit form in terms of position ($\phi = \phi(x)$) could also be found, through the Poisson equation.

\begin{figure*}
\begin{center}
\includegraphics{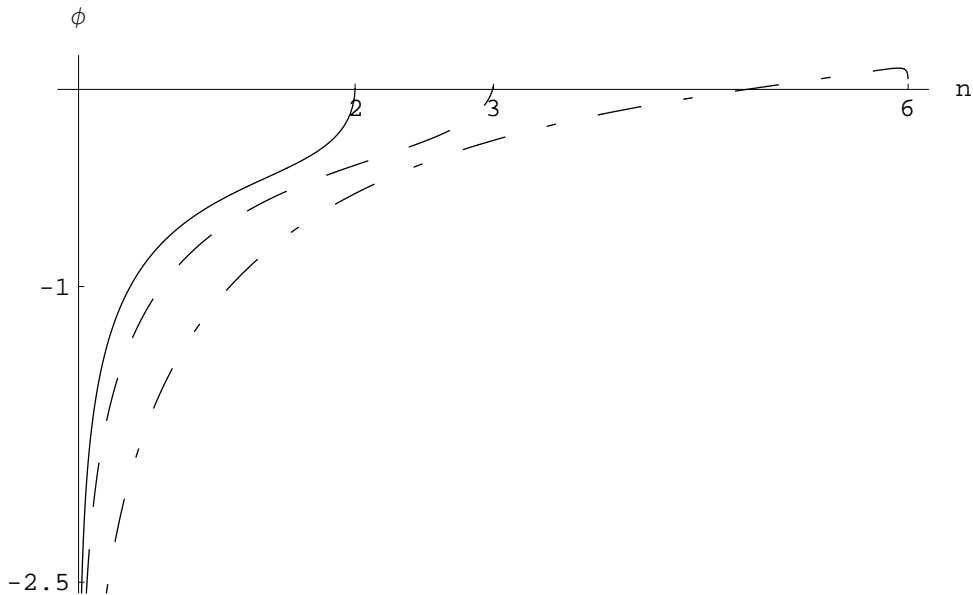}
\caption{the electrostatic potential as a function of the electron number density, for $e =m = \hbar = n_0 = \omega_p  = B_0 = 1$ and $C_0 = 1$ (upper curve, dot-dashed), $C_0 = 2$ (mid curve, dashed) and $C_0 = 3$ (lower curve, solid line).}
\label{figure7}
\end{center}
\end{figure*}

In the Wigner-Poisson case, the calculations becomes too involved if higher-order quantum corrections are added in the right-hand side of Eq. (\ref{wpq}). However, the stationary solution found are non perturbative, in the sense that they do not have, as a starting point, a prescribed zeroth-order Wigner function, like a Maxwell-Boltzmann or Fermi-Dirac equilibrium, as 
in Refs. \cite{luque}--\cite{wig}. In other words, except for the substitution $B \rightarrow B_0$ in the right-hand side of Eq. (\ref{eq20}), what have been found here is an exact solution for an approximate model, namely the Wigner-Poisson system up to the first-order quantum term. 

\section{Conclusion}
In this paper, we have presented explicit nonlinear solutions for the stationary Wigner and 
Wigner-Poisson equations. The solutions are inspired by the Bernstein-Greene-Kruskal modes of the classical Vlasov-Poisson plasma. However, unlike for the Bernstein-Greene-Kruskal modes, the equilibrium distribution function here depends not only on the energy, but also on the position. In addition,  there is no need for the eigen-functions or the characterization of the quantum statistical mixture of the system, with no reference to the Schr\"odinger equation. Our new solutions are in the form shown in Eqs. (\ref{eq8}) or (\ref{eq15}), which disentangle the equilibrium Wigner equation, thanks to the particular exponential dependence on the energy. It would be relevant to obtain another classes of explicit solutions, in order to derive a better understanding of the relation between the classical constants of motion and the equilibrium states of the Wigner and Wigner-Poisson equations.  Finally, there are no known necessary {\it and sufficient} criteria for a faithful Wigner function in the phase space. The solutions derived have 
been checked against the necessary conditions  Eqs. (\ref{c1})--(\ref{c2}),  for specific parameters only. 

\vskip .5cm
{\bf Acknowledgments}
\vskip .5cm

One of the authors (F.H.) thanks the Alexander von Humboldt Foundation for financial support.

\renewcommand{\baselinestretch}{2}

\end{document}